\documentclass[reprint,longbibliography,
 amsmath,amssymb,superscriptaddress]{revtex4-2}
\usepackage{graphicx}
\usepackage{dcolumn}
\usepackage{bm}
\usepackage[colorlinks=true, citecolor=blue, linkcolor=blue, urlcolor=blue]{hyperref}
\usepackage{xcolor}
\usepackage{amsmath, amssymb, amsfonts}
\usepackage{mathtools}
\usepackage{subfigure}
\usepackage{mathrsfs}
\usepackage{sidecap}	
\usepackage{verbatim}

\definecolor{purple1}{rgb}{128,0,128}

\usepackage{bigints}
\newcommand{\bea}{\begin{eqnarray}}
\newcommand{\ea}{\end{eqnarray}}

\definecolor{darkpastelgreen}{rgb}{0.01, 0.75, 0.24}

\def\d{\mathrm{d}}

\begin{document}

\title{
Quantum Brownian motion of a scalar particle in the presence of a partially reflecting boundary} 

\author{Caio C. \surname{Holanda Ribeiro}}
\email{caiocesarribeiro@alumni.usp.br}
\affiliation{International Center of Physics, Institute of Physics, University of Brasilia, 70297-400 Brasilia, Federal District, Brazil} 
\author{Vitorio A. \surname{De Lorenci}}
\email{delorenci@unifei.edu.br}
\affiliation{Instituto de F\'{\i}sica e Qu\'{\i}mica, Universidade Federal de Itajub\'a, Itajub\'a, Minas Gerais 37500-903, Brasil}
\affiliation{{${\cal G}\mathbb{R}\varepsilon\mathbb{C}{\cal
O}$}---Institut d'Astrophysique de Paris, CNRS \& Sorbonne
Universit\'e, UMR 7095 98 bis Boulevard Arago, 75014 Paris,
France}

\date\today

\begin{abstract}
The second quantization of a real massless scalar field in the presence of a material medium described by a Drude-like susceptibility is here examined in a 1+1 dimensional model. The modified vacuum fluctuations of this field imprint divergence-free velocity dispersions on a scalar-charged particle, thus elucidating that the origin of divergences that appear in previous treatments  is the assumption of idealized boundary conditions. Among the findings there is an oscillation on the dispersion curves caused by the effective mass of the field modes inside the dispersive medium. Additionally, it is found that the effects of the medium on the particle are delayed when compared to the perfect mirror limit, a phenomenon attributed to the imperfect reflection of field modes on the mirror. Although the study focus on a scalar field, the findings are valuable for understanding  models based on electromagnetic interaction. 
\end{abstract}

\maketitle

\section{Introduction}
The quantization of a scalar field $\phi(t,x,y,z)$ near a perfectly reflecting boundary (a mirror) leads to a  Hadamard function $\langle \{\phi(t,{\bf x})\phi(t',{\bf x'})\}\rangle$ \cite{Birrel&Davies} that diverges everywhere in the limit of point-coincidence. This two-point function can be renormalized by subtracting the Minkowski vacuum contribution, rendering a result that in the limit of point-coincidence is regular everywhere except on the mirror.
In principle, physical predictions can be derived from this renormalized propagator $\langle \{\phi(t,{\bf x})\phi(t',{\bf x'})\}\rangle_{\rm Ren}$. 
In the field quantization procedure, the presence of a mirror is stated by means of a boundary condition. For instance, if the mirror is placed at $x=0$, Dirichlet boundary condition will impose that the field must be zero on it, i.e., $\phi(t,x=0,y,z)=0$. This condition expresses the hypothesis of perfectness of the reflecting boundary. As a consequence, the Hadamard function will vanish on the mirror and renormalization procedure will not be able to remove the fundamental divergence at $x=0$.
Hence, this reminiscent divergence resists renormalization, and has its utmost origin in such mathematical assumption of a perfect mirror.
%

There is another divergence that appears in some observables which can also be associated with the above mentioned idealization. For instance, when a charged particle is placed at a distance $x$ from the boundary, dispersions of its position and velocity will be ill defined on the wall, as expected, but also after an interaction time $\tau = 2x$ that corresponds to a round trip of a light signal between the particle and the mirror \cite{ford2004,Lorenci2014}.
Notice that as soon as the particle is placed at its initial position $x$, it will immediately be under action of the modified (by the presence of the mirror) vacuum fluctuations of the background field. Its presence will affect the charge distribution of the mirror only after a time $\tau =x$, even though the consequences of this effect are usually neglected by assuming a test particle with a negligible charge. Finally, only after a time $\tau=2x$ the particle measures the presence of the mirror by means of its own reflex.
Recap that in the canonical quantization procedure the scalar field is treated as an operator that is expanded in a complete set of normal modes, each mode being a harmonic oscillator of a certain frequency. These infinite frequency-modes are all reflected by the mirror, whose perfect reflectivity is again the source of the divergence at $\tau = 2x$. 

In order to regularize these divergences, but still keeping the mathematical convenience of imposing Dirichlet boundary condition, the introduction of sample functions is usually adopted in the literature. Sample functions \cite{Ford2021,Barton_1991} can be used to blur the position of the mirror \cite{Lorenci2014}, or to implement a smooth mechanism of turning-on and off the interaction \cite{Lorenci2019}, among other possibilities. The use of smooth sample functions is certainly a step forward bringing more reality to the behavior of the system, but it is still a mathematical artifact that miss information about the very behavior of the system under examination. 

A more realistic account should avoid the over-idealization brought by the assumption of a perfect mirror. A real reflecting boundary is an extensive material whose reflectivity depends on the frequency of the incident wave. In this sense, the previous boundary condition must be exchanged by the information about the optical properties of the medium, which is usually described by means of its frequency-domain susceptibility tensor ${\bf\chi}_\omega(x)$. In other words, in the presence of a real reflecting material the quantization scheme takes into account that the field modes can be partially reflected and partially transmitted through it, depending on the magnitude of their frequencies.

In this paper a toy model based on a real massless scalar field in 1+1 dimensional spacetime half-filled with a material medium characterized by a frequency-dependent ``susceptibility'' function is investigated. A  test particle of mass $m$ and scalar charge $g$ is left to interact with the modified vacuum fluctuations of the background field whose quantization is implemented by assuming that the ``optical'' properties of the  medium is described by a Drude-like susceptibility. Dispersions of the particle velocity are thus calculated and compared with solutions that are obtained by using idealized models. The results are naturally free of the divergences above discussed. Furthermore, new features on the behavior of the dispersions are unveiled, highlighting the displacement of its minimum value from $\tau=2x$ and the presence of oscillations that are linked to the effective mass of the field quanta due to the presence of the material medium. 

In the next section some basic features about the quantum Brownian motion of a charged scalar particular interacting with a background field are presented.
The field equations are discussed in Sec.~\ref{fieldequation}, where a model for the susceptibility of the material half-space is discussed. The main results on the quantization of the system is thus presented in Sec.~\ref{fieldquantization} where the expression for the renormalized propagator is derived. The results are thus used in the Sec.~\ref{velocity}, where the dispersion of the particle velocity is calculated and numerically examined. A comparison with previous results is provided. In particular, it is shown that well known divergences associated with the idealized boundary condition do not appear in this formulation. Final remarks and conclusions are presented in Sec.~\ref{conclusions}. Full details on the quantization of the system are developed in the appendix~\ref{appendixa}. Finally, in appendix~\ref{appendixb} the evolution of a Gaussian wave packet is examined as it is reflected by the boundary of the material medium at $x=0$. 

Units are such that $\hbar=c=1$.


\section{Quantum Brownian motion}

Let $\phi=\phi(t,x)$ denote a relativistic massless real scalar field in $1+1$ dimensions and suppose that a non-relativistic scalar test particle can be used to probe the background field $\phi$. Specifically, if $m$ and $g$ denote the particle's mass and scalar charge, respectively, in the non-relativistic regime its interaction with $\phi$ is modeled by the Newtonian law \cite{Lorenci2014}
\begin{equation}
m\frac{\d v}{\d t}=-g\frac{\partial \phi}{\partial x},\label{vder}
\end{equation}
where $x=x(t)$ denotes the particle's position and $v=\d x/\d t$. If the particle is at rest at $t=0$, Eq.~\eqref{vder} can be integrated to obtain the particle velocity in general as
\begin{equation}
v(\tau)=-\frac{g}{m}\int_{0}^{\tau}\d t\frac{\partial \phi}{\partial x}(t,x(t)),\label{vel}
\end{equation} 
which is an integro-differential equation for the unknown $x(t)$. Following the discussion in \cite{Lorenci2014}, in the non-relativistic regime it can be assumed that the particle position $x(t)$ is approximately constant during the {\it measuring} time $\tau$, for which case, in a first approximation, the time-dependence of $\partial \phi/\partial x$ through $x(t)$ in Eq.~\eqref{vel} can be neglected \cite{Lorenci2014,Lorenci2019}.

Furthermore, if the field $\phi$ is a quantum field, the particle velocity $v$ given by Eq.~\eqref{vel} becomes an operator-valued distribution that can be used to study dispersions induced by quantum fluctuations of $\phi$. It is here assumed that $\phi$ is prepared in a vacuum state, for which case $\langle\phi\rangle=0$, but $\langle\phi^2\rangle\neq0$. Accordingly, it follows that $\langle v\rangle=0$, and thus measurements of the particle velocity are distributed around zero with variance $\langle(\Delta v)^2\rangle=\langle(v-\langle v\rangle)^2\rangle=\langle v^2\rangle$. This is the essence of the Quantum Brownian Motion.  

\section{Field equation in the presence of an imperfect dieletric}
\label{fieldequation}

 In order to write down a general equation of motion for $\phi$, let us start from the usual Klein-Gordon equation $(\partial^2_t-\partial_x^2)\phi=0$, and perform the Fourier decomposition
\begin{equation}
\phi(t,x)=\frac{1}{\sqrt{2\pi}}\int_{-\infty}^{\infty}\d\omega e^{-i\omega t}\phi_{\omega}(x),
\end{equation} 
such that $\phi_\omega^*=\phi_{-\omega}$ and $(-\omega^2-\partial_x^2)\phi_\omega=0$. A material medium is modeled by adding to the latter equation  a frequency-dependent 
 ``scalar field permittivity,''
\begin{equation}
[-\omega^2\epsilon_\omega(x)-\partial_x^2]\phi_\omega=0,\label{eqfourier}
\end{equation} 
under the physical requirement that $\chi_\omega(x):=\epsilon_\omega-1\rightarrow 0$ as $|\omega|\rightarrow\infty$, where $\chi_\omega(x)$ plays the role of a field-susceptibility of the material medium, in equivalence with the notation used in the case of the electromagnetic interaction. This allows modeling a sort of dielectric medium for the scalar field that is transparent to high-frequency field modes.

The quantity $\chi_\omega(x)$ cannot be an arbitrary function of $\omega$, as the underlying theory must respect causality. Indeed, by taking the inverse Fourier transform of Eq.~\eqref{eqfourier} we obtain that \cite{Fardin}
\begin{equation}
(\partial_t^2-\partial^2_x)\phi(t,x)+\partial_t\int_{-\infty}^{\infty} \d t'\chi(t',x)\partial_t\phi(t-t',x)=0,\label{fieldeq1}
\end{equation}
where the linear response function is
\begin{equation}
\chi(t,x)=\frac{1}{2\pi}\int_{-\infty}^{\infty}\d\omega\,  \chi_{\omega}(x)e^{-i\omega t}.
\end{equation}
Thus, it follows from Eq.~\eqref{fieldeq1} that causality is ensured as long as $\chi(t,x)=0$ for $t<0$, in which case
\begin{equation}
\chi_{\omega}(x)=\int_0^{\infty}\d t \, \chi(t,x)e^{i\omega t},
\end{equation}
viewed as a function of $\omega$ is analytical in the upper half complex plane. In particular, it satisfies the Kramers-Kronig relations \cite{Loudon1995}.

In what follows, only theories such that $\chi_{\omega}(x)$ is a Drude-type susceptibility and factorizes as $\chi_{\omega}(x)=\Theta(-x)f(\omega)$ are considered, where
\begin{equation}
f(\omega)=\frac{i\sigma_0}{\omega}\frac{1}{1-ib\omega},\label{dmodel}
\end{equation}
$\Theta(x)$ is the unit step function: $\Theta(x)=1$ if $x\geq0$ and 0 otherwise. In particular, $\chi_{\omega}$ analytic in the upper half plane implies that $b>0$, $\sigma_0>0$, which in the present case is sufficient to guarantee we have an absorbing (stable) dieletric, i.e., $\mbox{Im}(\chi_{\omega})>0$ for $\omega>0$ \cite{Shepherd}. We note that $\sigma_0\rightarrow\infty$ enforces that $\phi=0$ is the only normalizable solution of Eq.~\eqref{fieldeq1} for $x<0$. Thus this model allows one to study the scalar field dynamics in the presence of half space filled by a dispersive dielectric that contains the dispersionless perfect mirror at $x=0$ as limiting case.

\section{Field quantization}
\label{fieldquantization}

In this section a general expression for the Wightman function $\langle\phi(t,x)\phi(t',x')\rangle$ for the dielectric model of Eq.~\eqref{dmodel} is presented. Note that canonical quantization is not possible for this theory, as causality, through the Kramers-Kronig relations, implies that $\chi_\omega$ is always a complex function, and so the field equation Eq.~\eqref{eqfourier} is also complex. In physical terms, this system is necessarily dissipative. In what follows, the method of Langevin operators \cite{Loudon1995,Loudon2} is adopted in order to obtain a quantum field expansion for $\phi$ valid for the empty space region ($x>0$).

In the absence of the medium, it is straightforward to show that $\phi$ for $x>0$ can be expanded as
\begin{equation}
\phi=\frac{1}{\sqrt{2\pi}}\int_0^\infty\d\omega\frac{e^{-i\omega t}}{\sqrt{2\omega}} \left(a_{L,\omega}e^{-i\omega x}+a_{R,\omega}e^{i\omega x}\right)+\mbox{H.c.},\label{qfe}
\end{equation}  
where $L$, $R$ stand for leftwards, rightwards, respectively, and indicate if the field mode is propagating towards or away from $x=0$. Moreover, the canonical commutation relation reads $[a^{}_{L,\omega},a^\dagger_{L,\omega'}]=[a^{}_{R,\omega},a^\dagger_{R,\omega'}]=\delta(\omega-\omega')$, and all other commutators vanish. In particular, there is no correlation between leftwards and rightwards propagating modes: $[a^{}_{L,\omega},a^\dagger_{R,\omega'}]=0$.

When the dispersive medium is present,  the quantum field expansion can be written in the same form of Eq.~\eqref{qfe} (see the appendix \ref{appendixa} for details), with the difference that in this scenario leftwards and rightwards propagating waves are correlated, i.e.,
\begin{equation}
[a^{}_{R,\omega},a^\dagger_{L,\omega'}]=R_{\omega}\delta(\omega-\omega'),
\end{equation} 
where $R_{\omega}$ is the reflection coefficient obtained by solving the scattering problem for waves of frequency $\omega$ reaching the material medium from the empty space region,
\begin{equation}
R_{\omega}=\frac{1-n_{\omega}}{1+n_{\omega}},
\end{equation}
with $n_{\omega}=\sqrt{1+f(\omega)}$ being the ``index of refraction''. Furthermore, $R_{\omega}^*=R_{-\omega}$ for all real $\omega$, and $R_{\omega}$, viewed as a function of $\omega$, is analytic and bounded in the upper half place. From this property it follows that for $x,x'>0$
\begin{equation}
[\phi(t,x),\partial_t\phi(t,x')]=i\delta(x-x')+\frac{i}{2\pi}\int_{-\infty}^{\infty}\d\omega R_{\omega}e^{i\omega\hat{\Delta}x},
\end{equation}
$\hat{\Delta}x:=x+x'$, and the integral on the right hand side of the above equation is zero, as seen by closing the integration contour from the above. 

Finally, with the creation and annihilation operators identified, the vacuum state $|0\rangle$ is defined as $a_{L,\omega}|0\rangle=a_{R,\omega}|0\rangle=0$ for all $\omega>0$, from which the Wightman function is shown to acquire the form $\langle\phi(t,x)\phi(t',x')\rangle=\langle\phi(t,x)\phi(t',x')\rangle_{0}+\langle\phi(t,x)\phi(t',x')\rangle_{\rm Ren}$, where $\langle\phi(t,x)\phi(t',x')\rangle_{0}$ is the empty space two-point function \cite{Lorenci2019} and
%
%
\begin{equation}
\langle\phi(t,x)\phi(t',x')\rangle_{\rm Ren} = \frac{1}{4\pi}\int_{-\infty}^{\infty}\d k\frac{e^{-i\omega \Delta t}}{\omega}R_{k}e^{ik\hat{\Delta} x},\label{wightman}
\end{equation}
where now $\omega =|k|$. The subindex ``$\rm Ren$'' denotes the renormalized two-point function, i.e., with the empty space contribution subtracted. Equation \eqref{wightman} is valid for all absorbing media for which $\chi_{\omega}(x)=\Theta(-x)f(\omega)$. Within the model of Eq.~\eqref{dmodel} it follows that as $\sigma_0\rightarrow\infty$, $R_{k}\rightarrow -1$, and the renormalized Wightman function for the perfect (Dirichlet) mirror defined by $\phi(t,0)=0$ is found \cite{Lorenci2014,Lorenci2019}, as expected. 

Note that in the perfect mirror limit, the reflection coefficient is \emph{constant} for all frequencies, showing that all modes are affected by the mirror in the same manner. In contrast, for the material medium under consideration, the reflection coefficient behaves as 
\begin{equation}
R_{\omega}\rightarrow \frac{\sigma_0}{4b\omega^2},~\label{extrad}
\end{equation}
as $|\omega|\rightarrow \infty$, providing an extra decay in the correlation \eqref{wightman} for high energy modes. This theory, therefore, is naturally well-behaved in the UV sector. Note, however, that the unavoidable infrared divergence of two-dimensional massless field theories \cite{FULLING1987135} is present in the renormalized Wightman function, as it should.

\section{Velocity dispersions}
\label{velocity}

\begin{figure}[t]
\includegraphics[scale=0.55]{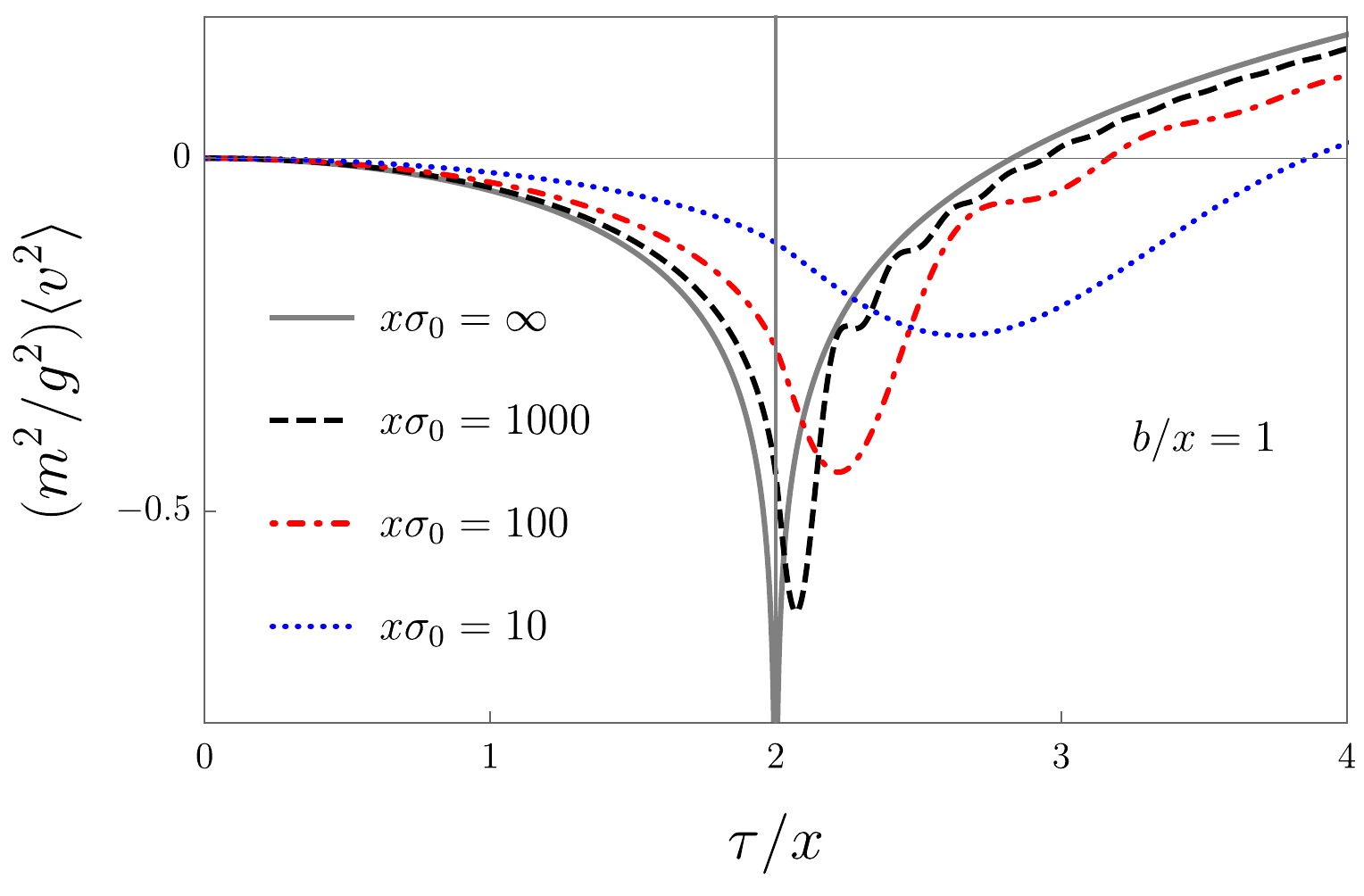}
\caption{Velocity dispersions for the susceptibility presented in Eq.~\eqref{dmodel}. We set $b/x=1$. Continuous gray line depicts the dispersions for the case of a Dirichlet mirror [Eq.~\eqref{dirichlet}]. Dashed, dot-dashed, and dotted curves show that the velocity dispersions are divergence-free when the dispersive nature of material media is taken into account. The valleys of the latter curves do not occur at $\tau=2x$ as happens  in the Dirichlet limit, a phenomenon linked to the increased penetration depth into the material of higher frequency modes. Also, we note the presence of oscillations after $\tau=2x$, suggestive of massive field behavior \cite{Lorenci2019}.}
\label{fig1}
\end{figure}

With the aid of the renormalized two-point function, it is straightforward to show that the velocity dispersion $\langle v^2\rangle$ assumes the form
\begin{equation}
\langle v^2\rangle=-\frac{g^2}{2\pi m^2}\int_{-\infty}^{\infty}\d k\frac{1-\cos\omega \tau}{\omega}R_{k}e^{2ik x}.\label{veldisp}
\end{equation}
%
%
%
%
%
In particular, in the perfect mirror limit, Eq.~\eqref{veldisp} can be exactly integrated to obtain \cite{Lorenci2014}
\begin{equation}
\langle v^2\rangle_{\rm Dirichlet}=\frac{g^2}{2\pi m^2}\ln\left|1-\frac{\tau^2}{4x^2}\right|,\label{dirichlet}
\end{equation}
from which the divergence at $\tau=2x$ is manifest. Now, because the latter occurs sharply at $\tau=2x$, it depends on how high frequency modes are scattered by the material, and it is not expected to occur for realistic scenarios where the effects of the high energy field modes are removed by  renormalizing the correlations, as shown in \cite{Lorenci2019}. For the dispersive medium under consideration, Eq.~\eqref{extrad} shows that the material transparency for high frequency modes ensure that the velocity dispersions modeled by Eq.~\eqref{veldisp} are divergence-free, as depicted in Fig.~\ref{fig1}.

Figure \ref{fig1} also reveals that for the perfect mirror limit the velocity dispersion presents a valley around $\tau=2x$ where the (logarithmic) divergence occurs. In contrast,    the global minima of all the curves obtained for the dispersive materials occur {\emph{after}} $\tau=2x$. This behavior is attributed to the longer time taken for wave packets sent towards the dielectric to be scattered back to the source in comparison to the perfect mirror setup. This phenomenon is illustrated in the appendix \ref{appendixb} for a Gaussian pulse.

Also, Fig.~\ref{fig1} unveils the existence of an oscillatory behavior after $\tau=2x$ which is absent in the Dirichlet mirror. A similar phenomenon was reported in \cite{Lorenci2019} for the velocity dispersions when the background field was massive. Here, although the background field is massless, the field dispersion relation inside the medium reads $\omega^2n_{\omega}^2=k^2$, and becomes $\omega^2=k^2+\sigma_0/b$ for $\omega\rightarrow\infty$, which is the typical dispersion relation of a massive field with mass $(\sigma_0/b)^{1/2}$. Accordingly, high frequency modes inside the medium behave effectively as massive modes, from which the observed oscillations are expected.

Finally, it is instructive to compare the divergence-free dispersion of Eq.~\eqref{veldisp} with the regularized dispersions found with the use of switching functions. The latter are calculated assuming a Dirichlet mirror in a scenario where the scalar particle starts to experience the background field state in a continuous manner, such that Eq.~\eqref{vel} can be amended as
\begin{equation}
v(\tau)=-\frac{g}{m}\int_{-\infty}^{\infty}\d tF_{\tau,\tau_s}(t)\frac{\partial \phi}{\partial x}(t,x),
\end{equation}
where $F_{\tau,\tau_s}$ is normalized as $\int_{-\infty}^{\infty}\d tF_{\tau,\tau_s}(t)=\tau$ and $\tau_s\rightarrow 0^{+}$ implies $F_{\tau,\tau_s}(t)\rightarrow \Theta(t)\Theta(\tau-t)$. The parameter $\tau_s$ is then identified as the switching time, and a convenient choice for a switching is $\pi F_{\tau,\tau_s}(t)=\arctan(t/\tau_s)+\arctan[(\tau-t)/\tau_s]$, such that the velocity dispersions can be exactly integrated in the limit of the Dirichlet mirror:
\begin{align}
&\langle v^2\rangle_{\rm Dirichlet,\tau_s}=\nonumber\\
&\frac{g^2}{4\pi m^2}\ln\left[\frac{(\tau^2-4x^2)^2+8(\tau^2+4x^2)\tau_s^2+16\tau_s^4}{16(x^2+\tau_s^2)^2}\right].\label{veldisp2}
\end{align}
Thus, when $\tau_s\rightarrow 0$ the sudden limit dispersion of Eq.~\eqref{dirichlet} is found, and Eq.~\eqref{veldisp2} is a regular function of $\tau$ for all $\tau_s>0$. 
\begin{figure}[t]
\includegraphics[scale=0.55]{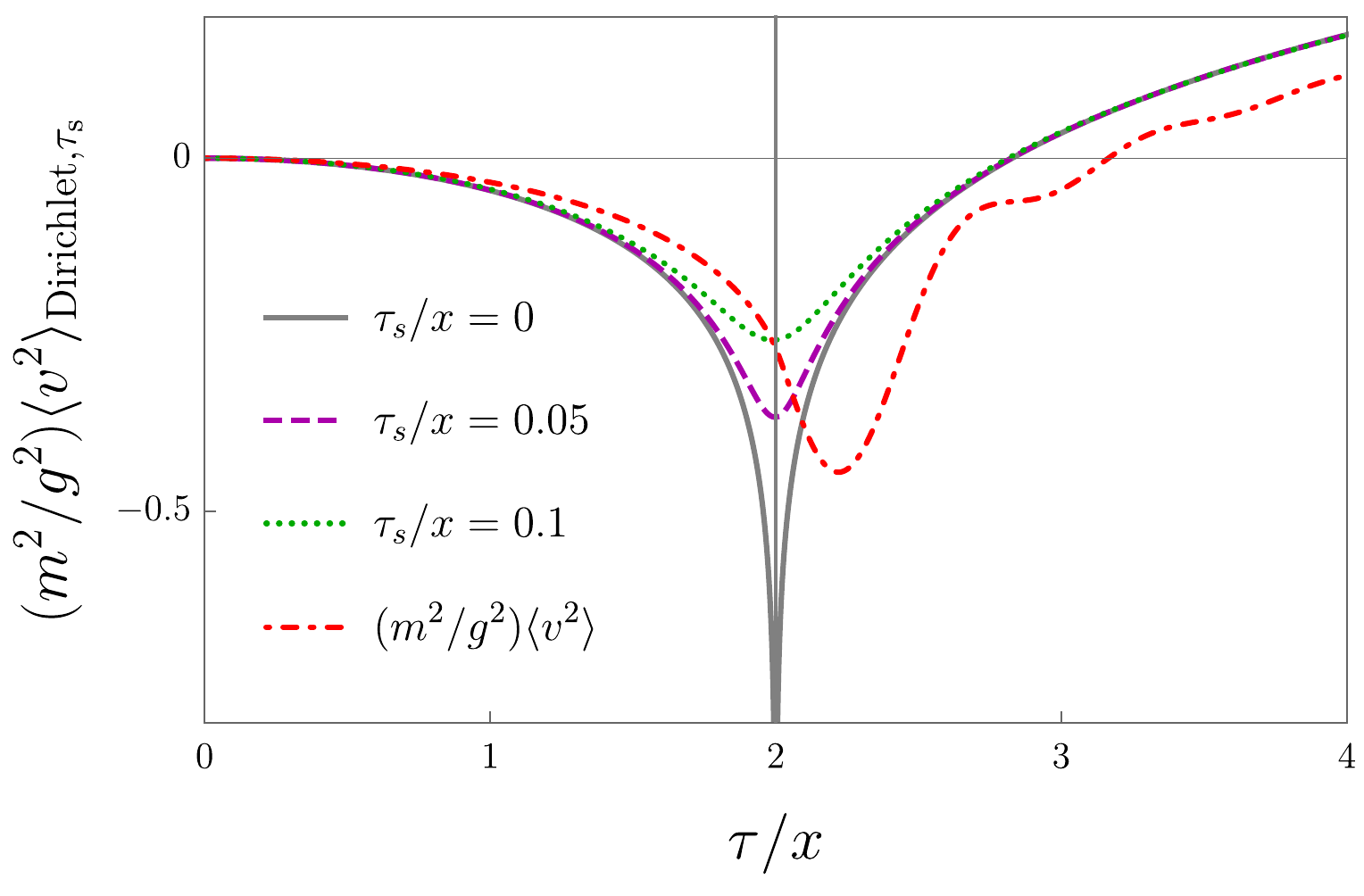}
\caption{Velocity dispersions for a Dirichlet mirror regularized by the switching function. Continuous gray line is the sudden regime $\tau_s=0$. Dashed and dotted curves are obtained for finite $\tau_s$, from which we see that the effect of the switching occurs only about the divergence at $\tau=2x$. Dot-dashed curve corresponds to the dot-dashed curve of Fig.~\ref{fig1}, and it is here for the sake of comparison.}
\label{fig2}
\end{figure}  
Figure \ref{fig2} shows that the effect of the switching function is to tame the divergent dispersions only around $\tau=2x$, allowing one to obtain estimates for the effect near $\tau=2x$ from exact expressions. However, as seen from the curves in Fig.~\ref{fig1} and the dot-dashed curve in Fig.~\ref{fig2}, the regularization obtained from the switching is not capable of capturing most of the dispersive material implications, as these cannot be extracted from  the perfect mirror limit in any meaningful way.

\section{Final Remarks}
\label{conclusions}
The renormalized correlation function defined by Eq.~(\ref{wightman}) generalizes previous results found in the literature. For instance, the Wightman function obtained when a perfect reflecting boundary condition is adopted is here found as the limiting case of $\sigma_0\to\infty$ ($R_{\omega}\to -1$), which is equivalent of assuming a material medium with infinite conductivity (a perfect conductor) in the electromagnetic case. 
It is worth noting that the dominant contribution of $R_{\omega}$ when $\omega\to 0$ is also $-1$, which coincides with the above discussed limit of a perfect conductor, showing that the Drude-like model under consideration enforces perfect reflection of low frequency modes and no reflection at all of higher frequency modes. Accordingly, renormalization with respect to Minkowski vacuum leads to well-behaved  theory in the UV sector.  Yet, the infrared divergence survives, which is a feature of massless quantum field theory in two-dimensional spacetime \cite{Birrel&Davies,FULLING1987135}.
This divergence has no consequences for the velocity dispersion, as it is removed under the action of the derivatives of the field correlations. 

Finally, although the scalar field model adopted here does not correspond to any known fundamental field, it is important to stress that analogous results and conclusions are found, for instance, for the electric monopole if only the quantum nature of electromagnetic waves propagating perpendicularly to a dielectric wall is considered as done in \cite{Loudon1995,Loudon2}.

%

%
%
%
%

\begin{acknowledgments}
This work was partially supported by the Brazilian research agency CNPq (Conselho Nacional de Desenvolvimento Cient\'{\i}fico e Tecnol\'ogico) under Grant No. 305272/2019-5.
\end{acknowledgments}

\appendix

\section{Field quantization}
\label{appendixa}

The quantization needed in order to deduce Eq.~\eqref{wightman} can be performed in a similar manner as was done, for instance, in Refs.~\cite{Loudon1995,Loudon2} for the electromagnetic field using Langevin operators. Specifically, we let the Hermitian operator $J$ be a Langevin source term such that the quantum field satisfies
\begin{align}
(\partial_t^2&-\partial^2_x)\phi(t,x)\nonumber\\
&+\partial_t\int_{-\infty}^{\infty} \d t'\chi(t',x)\partial_t\phi(t-t',x)
=J(t,x).\label{fieldeqA1}
\end{align}
%
%
%
For stationary configurations, $J$ also admits the Fourier decomposition
\begin{equation}
J(t,x)=\frac{1}{\sqrt{2\pi}}\int_{-\infty}^{\infty}\d\omega e^{-i\omega t}J_{\omega}(x),
\end{equation} 
with $J_{\omega}^{\dagger}=J^{}_{-\omega}$, and it is subjected to the commutation relations
\begin{equation}
[J^{}_{\omega}(x),J^\dagger_{\omega'}(x')]=2\omega^2\mbox{Im}(\epsilon_{\omega})\delta(\omega-\omega')\delta(x-x'),\label{ap4}
\end{equation}
and $[J_{\omega}(x),J_{\omega'}(x')]=0$. Accordingly, the field equation in frequency-domain reads
\begin{equation}
[-\omega^2\epsilon_{\omega}(x)-\partial_x^2]\phi_{\omega}(x)=J_{\omega}(x).\label{ap2}
\end{equation}
The most general solution of Eq.~\eqref{ap2} can be written as a sum of a particular solution $\phi_{\omega}^{p}$ plus solutions to the (homogeneous) sourceless equation $\phi_{\omega}^{h}$, and the method of Green functions is particularly useful to find particular solutions for the model under study. Indeed, if $G_{\omega}(x,x')$ satisfies
\begin{equation}
[-\omega^2\epsilon_{\omega}(x)-\partial_x^2]G_{\omega}(x,x')=\delta(x-x'),\label{apaprop}
\end{equation} 
then
\begin{equation}
\phi^{p}_{\omega}(x)=\int_{-\infty}^{\infty}\d x'G_{\omega}(x,x')J_{\omega}(x')
\end{equation}
solves Eq.~\eqref{fieldeqA1}. We note that $G_{\omega}(x,x')$ is not uniquely determined by Eq.~\eqref{apaprop} as boundary conditions must be specified in order to select a physical propagator. We work with the causal propagator for which $\omega$ has a small positive imaginary part included \cite{Birrel&Davies}.  Let us consider the Fourier transform $\tilde{G}_{\omega}(x,k)=\int_{-\infty}^{\infty}\d x'\exp(ikx')G_{\omega}(x,x')$, such that
\begin{equation}
[-\omega^2\epsilon_{\omega}(x)-\partial_x^2]\tilde{G}_{\omega}(x,k)=e^{ikx}.\label{ap3}
\end{equation}
For the case where $\epsilon_{\omega}(x)=1+\Theta(-x)f(\omega)$, the general solution of the above equation for $x\neq0$ reads
\begin{equation}
\tilde{G}_{\omega}(x,k)=\frac{e^{ikx}}{k^2-\omega^2\epsilon_{\omega}}+\Theta(x) A e^{i\omega x}+\Theta(-x)Be^{-i\omega n_{\omega}x},
\end{equation}
where we have defined the material index of refraction $n_{\omega}=[1+f(\omega)]^{1/2}$, and the coefficients $A$, $B$ are determined by imposing continuity of $\tilde{G}_{\omega}(x,k)$ and its derivative with respect to $x$ at $x=0$. The latter conditions are enforced by Eq.~\eqref{ap3}, and we find that
%
%
\begin{align}
A&=\frac{\omega(n_{\omega}-1)}{(k^2-\omega^2)(k-\omega n_\omega)},\nonumber\\
B&=\frac{\omega(n_{\omega}-1)}{(k^2-\omega^2n_{\omega}^2)(k+\omega)}.
\end{align} 
Thus, by taking the inverse transform $G_{\omega}(x,x')=(1/2\pi)\int\d k\exp(-ikx')\tilde{G}_{\omega}(x,k)$, we find that, for $x>0$ and $x'<0$,
\begin{equation}
G_{\omega}(x,x')=i\frac{e^{i\omega( x- n_{\omega}x')}}{\omega(1+n_{\omega})}.
\end{equation}
The above propagator describes waves at $x>0$ that were originated at $x' < 0$ (inside the medium). It vanishes when the limit of $n_{\omega}\to\infty$ is considered, as expected, and it and recovers the free space causal propagator when $n_{\omega}\to 1$. Therefore,
\begin{equation}
\phi_{\omega}^{p}(x)=\frac{i}{\omega}\frac{e^{i\omega x}}{1+n_{\omega}}\int_{-\infty}^{0}\d x'J_{\omega}(x')e^{-i\omega n_{\omega}x'}
\end{equation}
is the required particular solution for $x>0$.

Finally, the last ingredient for the quantization is to add relevant homogeneous solutions of Eq.~\eqref{ap2} to the field expansion. Because $\phi_{-\omega}=\phi_{\omega}^*$, we need only to focus on $\omega>0$. Let us note first that for $x<0$, the solutions of $[-\omega^2 n_{\omega}^2-\partial_{x}^2]\phi_{\omega}=0$ read $\exp(\pm i\omega n_{\omega}x)$, and because we are assuming $\mbox{Im}(\epsilon_{\omega})>0$ for $\omega>0$, only $\exp(- i\omega n_{\omega}x)$ is normalizable. As for $x>0$, the solutions are $\exp(\pm i\omega x)$, and both are normalizable, with $\exp(- i\omega x)$ corresponding to a plane wave propagating towards the dielectric. Thus, we find that
\begin{equation}
\sqrt{2\omega}\phi^{h}_{\omega}(x)=(e^{-i\omega x}+R_{\omega}e^{i\omega x})\Theta(x)+T_{\omega}e^{-i\omega n_{\omega}x}\Theta(-x),
\end{equation}
is the homogeneous solution we need. The matching constants $R_{\omega}$, $T_{\omega}$ read
\begin{align}
R_{\omega}&=\frac{1-n_{\omega}}{1+n_{\omega}},\\
T_{\omega}&=\frac{2n_{\omega}}{1+n_{\omega}},
\end{align}
and the normalization $\sqrt{2\omega}$ is obtained by the empty-space field quantization far away from the dieletric $x\rightarrow \infty$. Therefore, the quantum field expansion for $x>0$ reads
\begin{equation}
\phi(t,x)=\frac{1}{\sqrt{2\pi}}\int_{0}^{\infty}\d \omega e^{-i\omega t}[a_{L,\omega}\phi_{\omega}^{h}(x)+\phi_{\omega}^{p}(x)]+\mbox{H.c.},
\end{equation}
and the operator $a_{L,\omega}$ commutes with $J_{\omega'}$, $J^\dagger_{\omega'}$. By separating the distinct components into leftwards and rightwards propagating waves, we obtain
\begin{equation}
\phi=\frac{1}{\sqrt{2\pi}}\int_{0}^{\infty}\frac{\d \omega}{\sqrt{2\omega}} e^{-i\omega t}[a_{L,\omega}e^{-i\omega x}+a_{R,\omega}e^{i\omega x}]+\mbox{H.c.},
\end{equation}
where
\begin{equation}
a_{R,\omega}=R_{\omega}a_{L,\omega}+\frac{i\sqrt{2/\omega}}{1+n_{\omega}}\int_{-\infty}^{0}\d x'J_{\omega}(x')e^{-i\omega n_{\omega}x'}.
\end{equation}
Thus, it follows from Eq.~\eqref{ap4} that $[a_{R,\omega},a^\dagger_{R,\omega'}]=\delta(\omega-\omega')$, and this concludes the quantization. 

\section{Scattering of wave signals sent towards the material medium}
\label{appendixb}

This appendix presents an integral formula for the back-scattering of wave packets sent from the vacuum sector towards the material medium at $x=0$. In general, the formulation of initial value problems for the theory under consideration is naturally convoluted by memory effects, and in order to avoid unnecessary complications it is here assumed that signals are always originated outside the material medium, where the theory is local. Specifically,  suppose $\phi_{0}=\phi_{0}(t,x)$ is any given wave packet such that $\phi_0=0$ for $x\leq0$ and $t\leq0$, i.e., the wave packet does not meet the material before $t=0$.
The causal development of $\phi_0$ to later times is then ruled by the Eq.~\eqref{fieldeq1}, with $\phi_0$ furnishing the Cauchy data at $t=0$. 

A convenient way of determining the evolution of $\phi_0$ to later times is to calculate the auxiliary field $\phi$, solution of 
\begin{align}
&(\partial_t^2-\partial^2_x)\phi(t,x)+\partial_t\int_{-\infty}^{\infty} \d t'\chi(t',x)\partial_t\phi(t-t',x)\nonumber\\
&=\phi_0(0,x)\partial_t\delta(t)+[\partial_t\phi_0(t,x)]\delta(t),\label{apb1}
\end{align}
under the condition that $\phi=0$ for $t<0$. Then 
it is straightforward to see that $\phi$ coincides with the causal development of $\phi_0$ for $t>0$, i.e., $\phi(0,x)=\phi_0(0,x)$ and  $\partial_t\phi(0,x)=\partial_t\phi_0(0,x)$.

\begin{figure}[t]
\includegraphics[scale=0.80]{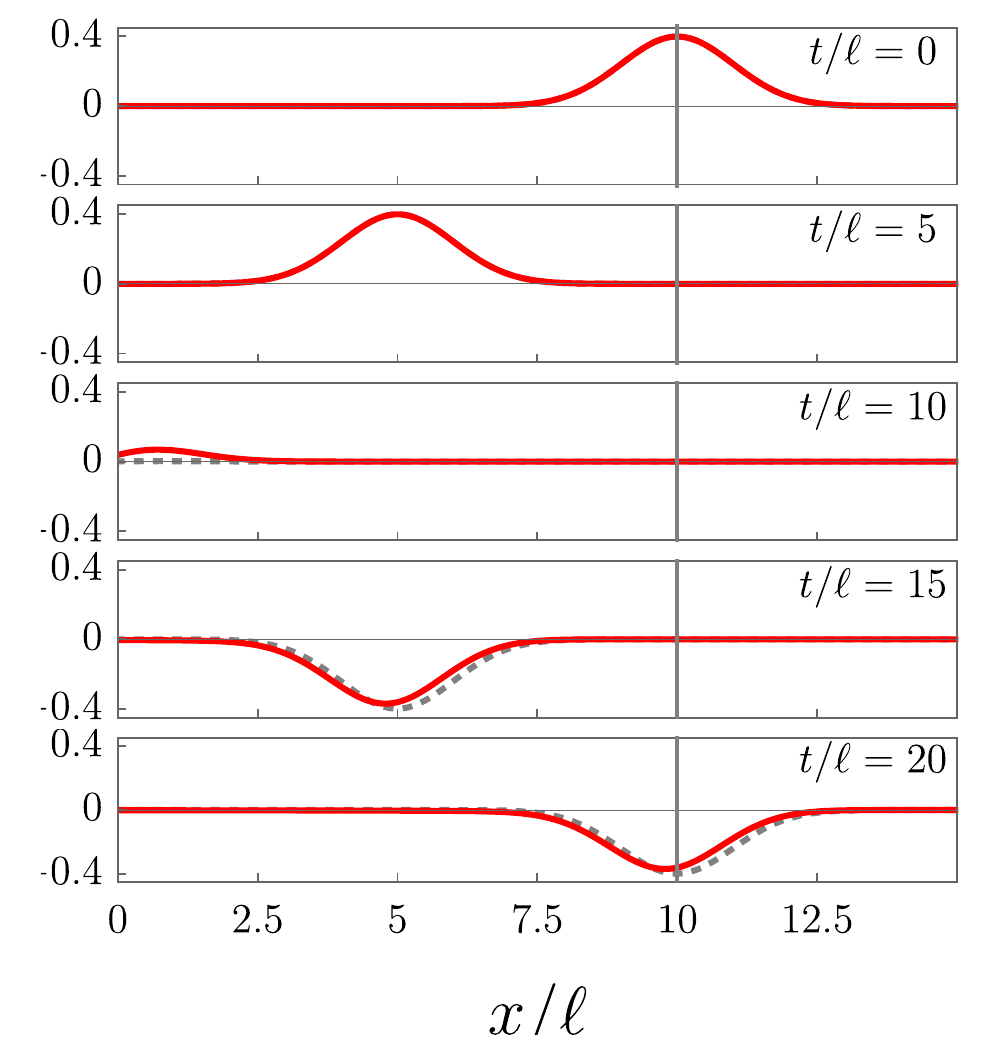}
\caption{Scattering of a Gaussian wave packet sent from the vacuum sector at the material medium at $x/\ell=0$. At $t/\ell=0$ the pulse is centered at $x_0/\ell=10$, and it is coherently travelling leftwards with velocity $1$. Dashed curve shows the back-scattering obtained for a perfect mirror, for which $\ell\sigma_0=\infty$, whereas the continuous curve corresponds to an imperfect mirror with $\ell\sigma_0=100$. Also, $b/\ell=1$. The figure shows that in the presence of a dispersive material, the pulse takes a longer time to return to its source.}
\label{fig3}
\end{figure}

Equation \eqref{apb1} can be solved by means of the causal propagator $G(t,x;t',x')$ solution of
\begin{align}
&(\partial_t^2-\partial^2_x)G(t,x;t',x')\nonumber\\
&+\partial_t\int_{-\infty}^{\infty} \d \eta\chi(\eta,x)\partial_tG(t-\eta,x;t',x')\nonumber\\
&=\delta(t-t')\delta(x-x'),\label{apb2}
\end{align}
such that
\begin{align}
\phi(t,x)=\int\d t'\d x' G(t,x;&t',x')\{\phi_0(0,x')\partial_{t'}\delta(t')\nonumber\\
&+[\partial_{\eta}\phi_0(\eta,x')|_{\eta=0}]\delta(t')\}.
\end{align}
$G(t,x;t',x')$ can be found with the representation
\begin{equation}
    G(t,x;t',x')=\frac{1}{2\pi}\int_{-\infty}^{\infty}\d\omega e^{-i\omega\Delta t}G_{\omega}(x,x'),
\end{equation}
where $G_{\omega}(x,x')$ is solution of Eq.~\eqref{apaprop}. Thus, from the results presented in appendix \ref{appendixa}, it follows that
\begin{equation}
    G_{\omega}(x,x')=\frac{i}{2\omega}\left(e^{i\omega|\Delta x|}+R_{\omega}e^{i\omega\hat{\Delta}x}\right),
\end{equation}
for $x,x'>0$, and
\begin{align}
    G(t,x;t'&,x')=G_0(t,x;t',x')\nonumber\\
    &-\frac{1}{2\pi}\int_0^{\infty}\frac{\d k}{k}\mbox{Im}\left[R_k e^{-ik(\Delta t-\hat{\Delta} x)}\right],
\end{align}
where $G_0(t,x;t',x')$ is the empty space causal propagator. Finally,
\begin{align}
    \phi(t,&x)=\phi_{0}(t,x)-\frac{1}{2\pi}\int_0^{\infty}\frac{\d k}{k}\mbox{Im}\Bigg\{R_k e^{-ik(t-x)}\times\nonumber\\
    &\int_{-\infty}^{\infty}\d x'e^{ikx'}[\partial_{t'}\phi_0(t',x')|_{t'=0}-ik\phi_0(0,x')]\Bigg\},
\end{align}
is the causal development of $\phi_0$ for $t,x>0$. Figure \ref{fig3} depicts $\phi$ for the Gaussian pulse $\phi(t,x)=\exp[-(t+x-x_0)^2/(2\ell^2)]/\sqrt{2\pi\ell^2}$ centered at $x_0/\ell=10$ at $t/\ell=0$.
%
%
%

\bibliography{RQB2}

%
%


\end{document}